# Validation of Coronal Mass Ejection Arrival-Time Forecasts by Magnetohydrodynamic Simulations based on Interplanetary Scintillation Observations

Kazumasa Iwai[*,1], k.iwai@isee.nagoya-u.ac.jp
Daikou Shiota[2,1], shiotad@nict.go.jp
Munetoshi Tokumaru[1], tokumaru@isee.nagoya-u.ac.jp
Ken'ichi Fujiki[1], fujiki@isee.nagoya-u.ac.jp
Mitsue Den[2], den@nict.go.jp
Yûki Kubo[2], kubo@nict.go.jp

1. Institute for Space-Earth Environmental Research, Nagoya University, Furo-cho, Chikusa-ku, Nagoya, 464-8601, Japan
2. National Institute of Information and Communications Technology, 4-2-1 Nukui-kita, Koganei, Tokyo 184-8795, Japan

[*] **Corresponding author: Kazumasa Iwai**

## Abstract

Coronal mass ejections (CMEs) cause various disturbances of the space environment; therefore, forecasting their arrival time is very important. However, forecasting accuracy is hindered by limited CME observations in interplanetary space. This study investigates the accuracy of CME arrival times at the Earth forecasted by three-dimensional (3D) magnetohydrodynamic (MHD) simulations based on interplanetary scintillation (IPS) observations. In this system, CMEs are approximated as spheromaks with various initial speeds. Ten MHD simulations with different CME initial speed are tested, and the density distributions derived from each simulation run are compared with IPS data observed by the Institute for Space-Earth Environmental Research (ISEE), Nagoya University. The CME arrival time of the simulation run that most closely agrees with the IPS data is selected as the forecasted time. We then validate the accuracy of this forecast using 12 halo CME events. The average absolute arrival-time error of the IPS-based MHD forecast is approximately 5.0 h, which is one of the most accurate predictions that ever been validated, whereas that of MHD simulations without IPS data, in which the initial CME speed is derived from white-light coronagraph images, is approximately 6.7 h. This suggests that the assimilation of IPS data into MHD simulations can improve the accuracy of CME arrival-time forecasts. The average predicted arrival times are earlier than the actual arrival times.

These early predictions may be due to overestimation of the magnetic field included in the spheromak and/or underestimation of the drag force from the background solar wind, the latter of which could be related to underestimation of CME size or background solar wind density.

## Keywords

interplanetary scintillation, coronal mass ejection, space weather forecasting, magnetohydrodynamics, data assimilation.

## Introduction

Coronal mass ejections (CMEs) are eruptions of magnetized plasma structures generated by solar eruptive phenomena such as flares. CMEs occasionally reach the Earth and cause various disturbances of the space environment, where a vast amount of social infrastructure is in operation (Zhang et al. 2007; Gopalswamy et al. 2018). Therefore, forecasting the arrival time of CMEs is an important topic in the field of space weather research.

Many studies have attempted to forecast CME arrival using empirical models (e.g. Gopalswamy et al. 2001), and global magnetohydrodynamic (MHD) simulations of the heliosphere such as ENLIL (Odstrcil 2003), SUSANOO (Shiota et al. 2014; Shiota and Kataoka 2016), and EUHFORIA (Pomoell and Poedts 2018). These simulations calculate the background solar wind from magnetic field data of the solar surface and empirical models of the solar wind speed (Wang and Sheeley 1990, Arge and Pizzo 2000, Arge et al. 2004). Then, the simulations approximate the CMEs as simple structures such as cones or spheromaks. The modeled CMEs are placed at the inner boundary, which is approximately a few tens of solar radii (Rs), and their propagation is simulated to 1 AU.

The accuracy of CME arrival-time forecasts has been statistically validated in previous studies (e.g., Riley et al. 2018; Wold et al. 2018), which suggest an error of more than 10 h for current MHD simulations. There are many possible causes of this arrival-time error. Ambiguity of the modeled CME parameters is the first candidate. Many studies used white-light coronagraph images, which are typically provided by the Large Angle and Spectrometric Coronagraph (LASCO: Brueckner et al. 1995) onboard the Solar and Heliospheric Observatory (SOHO), to detect CMEs and derive characteristics such as CME speed, width, and tilt angle for input into CME models (Wold et al. 2018 and references therein). However, these observations typically contain errors due to projection effects (Temmer et al. 2009). In fact, some studies have suggested that forecasting accuracy can be improved by deriving the input parameters from multiple spacecraft observations (Millward et al. 2013) or by using the heliospheric imager (HI) onboard Solar TErrestrial Relations Observatory (STEREO) satellites (Möstl et al. 2014). There are many other potential reasons for the observed arrival-time error, such as interaction between CMEs and background solar wind, accuracy of the simulated background solar wind, and CME-CME interactions during propagation (e.g., Möstl et al. 2014; Millward et al. 2013.; Lee et al. 2013).

Turbulence contained in the solar wind plasma can scatter radio emission from extra-galactic radio sources, which is known as interplanetary scintillation (IPS, Hewish et al. 1964). Rapidly propagating CMEs sweep the background solar wind, forming dense regions in front of the CMEs. These regions can significantly scatter radio emissions. Hence, IPS observations can be used to detect CMEs propagating in interplanetary space (e.g., Tokumaru et al. 2003; 2005; Manoharan 2006; Glyantsev et al. 2015; Johri and Manoharan 2016). IPS data are also used as additional information for the inner boundary of global MHD simulations (Jackson et al. 2015; Yu et al. 2015).

Iwai et al. (2019) developed a new CME forecasting system that combines MHD simulations (SUSANOO-CME; Shiota and Kataoka 2016) with IPS observations at 327 MHz. They used this system to investigate a CME generated by an X9.3 flare on September 6, 2017. After comparing IPS data with that estimated by multiple CME simulations under different initial conditions, they found that the CME simulation that best estimates the IPS observation can predict the CME arrival time on the Earth more accurately. Although IPS data are available only after CMEs reach interplanetary space and cause less lead-time for prediction, they appear to improve the accuracy of prediction.

The accuracy of arrival-time forecasts using real-time MHD simulations can be improved by incorporating IPS data because such data can observe CMEs in the interplanetary space. However, the accuracies of arrival-time forecasts using MHD simulations with and without IPS data under real-time operations have not been evaluated. Therefore, the purpose of this study is to evaluate the performance of the SUSANOO-CME model combined with IPS data for real-time forecasting using various CME events. To achieve this, we simulate the arrival times of 12 CME events using MHD simulations with and without IPS data, and compare their accuracy.

## Methods

**MHD simulation (SUSANOO-CME)**
In this study, we used a 3D global MHD simulation of the inner heliosphere called SUSANOO-CME. Details of this simulation system have been described in previous studies (Shiota et al. 2014; Shiota and Kataoka 2016; Iwai et al. 2019); therefore, we only summarize the basics of the system here. The simulation region is a spherical shaped inner heliosphere between 25 and 425 Rs, formed by a Yin-Yang grid (Kageyama and Sato 2004). At the inner radial boundary, the magnetic field, velocity, density, and temperature of the background solar wind are given by the potential field source surface (PFSS) approximation from the photospheric magnetic field and empirical models of the solar wind (Arge and Pizzo 2000; Hayashi et al. 2003).

A CME was approximated as a spheromak and placed on the inner boundary. The initial parameters of the spheromak were defined from the Geostationary Operational Environmental Satellite (GOES) and LASCO observational data to enable automatic operation in real-time. The location and onset time of the CME were approximated to those of the corresponding X-ray flare observed by GOES. We assumed that the magnetic flux included in the spheromak was proportional to the corresponding X-ray flux and set 3.0, 1.0, and 0.3 $\times 10^{21}\ Mx$ for X, M, and C class flares, respectively. The radial and angular widths of the spheromak were also defined from the X-ray flux as follows: radial width = 2, 3, and 4 Rs and angular width = 30, 60, and 90° for CMEs with C, M, and X class flares, respectively. The tilt angle of the spheromak was 90° for all CMEs.

In this study, we focus on the initial speed of the CMEs, which was employed as the only free parameter in the MHD simulations. For each CME event, we employed 10 simulations with different CME initial speed; the maximum speed derived from the Computer Aided CME Tracking software (CACTus: Robbrecht and Berghmans 2004; http://sidc.be/cactus/) using the LASCO data, the initial speed listed in the LASCO catalog (Yashiro et al. 2004; https://cdaw.gsfc.nasa.gov/CME_list/), and 8 other simulations with initial speeds that were faster or slower than the CME speed in the LASCO CME catalog in 100-km/s intervals.

The simulation period started two days before and ended four days after the CME onset. The simulation also contained CMEs observed within two days of the CME of interest in order to determine the effect of CME-CME interactions. The initial speeds of these CMEs were fixed to those listed in the LASCO CME catalog so they have no free parameters.

**IPS observations**
IPS observations were conducted by the Institute for Space-Earth Environmental Research (ISEE), Nagoya University. We used the g-value of IPS (Gapper et al. 1982) obtained from the Solar Wind Imaging Facility (SWIFT: Tokumaru et al 2011), which is one of the three large IPS radio telescopes of ISEE. This telescope observes 50–70 radio sources each day throughout the year at an observation frequency of 327 MHz, which is able to detect the IPS signatures of solar wind between 0.2 AU and 1.0 AU.

**IPS estimation using MHD simulations**
The IPS g-value was estimated from the 3D density distribution obtained from the SUSANOO-CME simulation, as described in Iwai et al. (2019). In this system, the radio scintillation of each radio source was calculated from the density distribution along the line of sight to the radio source assuming a weak scattering condition (Young 1971). We also assumed that the density fluctuation included in the solar wind was proportional to the density of the solar wind. Figure 1 shows an example of the estimated IPS g-value distribution on the sky plane. The high g-value region, which indicates a high-density region along the line of sight, has a loop-like distribution. This is because the fast propagating spheromak sweeps the background solar wind, forming a high-density region at the front of the spheromak (Iwai et al. 2019).

Once we set the range of initial parameters of the spheromak (e.g., speed, location, widths), we are able to restrict the possible location range of the spheromak at any given time. Therefore, we selected ranges for the time period, elongation, and position angle of the radio source, as indicated by the white circles and red lines in Figure 1. We selected the radio sources included in the partial torus surrounded by the red lines, and compared the observed and simulated g-value for a given time.

The IPS g-value describes the ratio between the typical IPS amplitude around the observation period and the IPS amplitude observed at a given moment. Therefore, if the front of the CME passes the line of sight to the radio source, the observed g-value should be larger than 1. We selected a radio source with g > 1 included in the partial torus, which may be able to detect the CME. Then, we derived the positional difference between the radio source and the g-value peak of the simulation along the same position angle. Then, we derived the average positional difference of all selected radio sources for each simulation run. The simulation run that estimated the IPS g-value most consistently, i.e., the simulation run with the smallest average positional difference, was deemed the most reliable simulation, and was selected as the solution of the IPS-based SUSANOO-CME simulation. Figure 2a displays the solar wind speed along the Sun-Earth line simulated in this study. Figure 2b displays the time variations of the solar wind speed at the position of the Earth. The CME arrival time of each simulation is defined using the onset of velocity enhancement in this study. Finally, the accuracy of the CME arrival time of each simulation was evaluated by comparing the spheromak arrival time at the Earth with the observed shock arrival time at the Earth using the OMNI data (https://omniweb.gsfc.nasa.gov/).

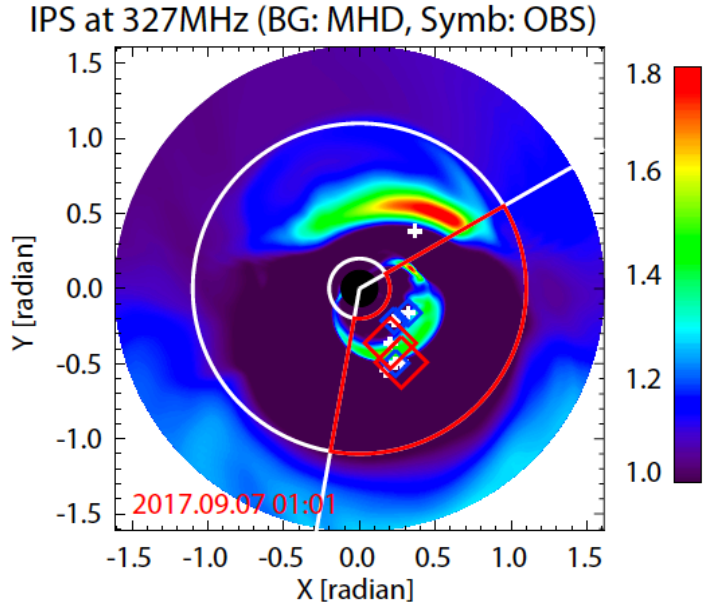

**Figure 1**. **Example distribution of the simulated IPS data.** IPS g-value estimated by the SUSANOO projected onto the sky-plane. White circles and red lines indicate elongation and position angle range of the radio source used for the forecasting, respectively. The symbols indicate the IPS g-value observed by SWIFT from 01:00 UT to 02:00 UT. Crosses indicate all observed radio sources and diamonds indicate radio sources with the following g-values: 2.0 < g (red), 1.5 < g < 2.0 (green), and 1.2 < g < 1.5 (blue).

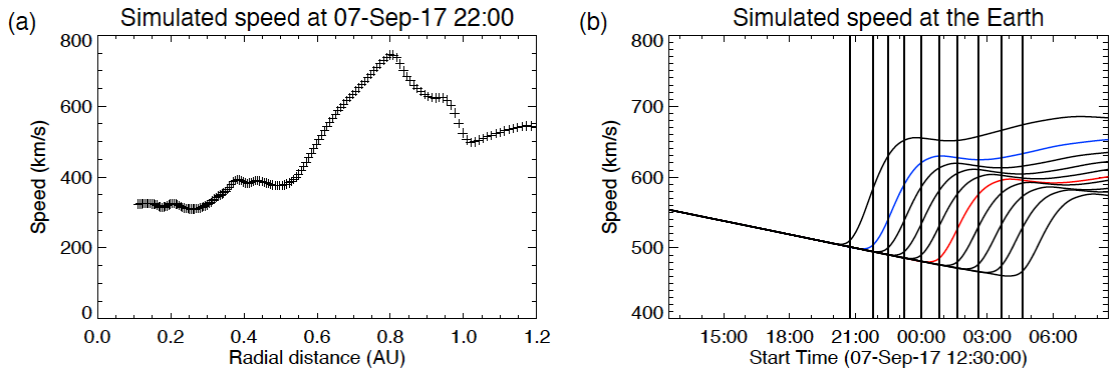

**Figure 2**. **Example of the simulated solar wind speed**. (a) An example of the simulated solar wind speed at each grid located along the Sun-Earth line (b) Solar wind speed at the Earth derived by simulation runs with CMEs with an initial speed derived from CACTus max speed (blue), the LASCO catalog (red), and eight CMEs with initial speeds that are faster or slower than the CME speed in the LASCO CME catalog with a 100-km/s interval (black). Vertical line indicates the arrival time of each CME at the Earth.

**Data set**

We selected CMEs for arrival-time validation according to the following selection criteria. The CMEs were halo CMEs observed between January 2013 and December 2018, with an initial speed in the LASCO CME catalog of greater than 1000 km/s. All CMEs were associated with M or X class flares, and corresponding interplanetary CMEs (ICMEs) were observed on the Earth. Both IPS observations of ISEE and white-light observations of LASCO were available during their propagation, and real-time simulation results for the arrival times were archived in the CME arrival-time scoreboard (Riley et al. 2018) of the Community Coordinated Modeling Center (CCMC). Our simulation could not forecast the arrival of some CMEs that occurred around the solar limb; therefore, these CMEs (that occurred from active regions located at a longitude of greater than 50 degree) were removed from the evaluation of arrival-time error. The 14 events that satisfy the criteria are listed in Table 1. It should be noted that despite the above criteria being satisfied, two CME events were removed owing to the obvious CME–CME interaction during propagation (January 30, 2014) and lack of IPS radio source along the possible trajectory of the CME (December 28, 2015). Therefore, 12 CME events were investigated in this study.

## Results

The forecasted arrival times for each CME are summarized in Table 1. In this table, the onset day, X-ray class and location of the corresponding flare, and arrival time of the observed CME are listed in columns 2–6. The simulated arrival times and the arrival-time errors of the simulation runs using CMEs with different initial speeds (CACTus maximum speed, LASCO initial speed, and the speed that best fits the IPS observations) are listed between columns 7–15. The real-time forecast result using the WSA-ENLIL-Cone model are archived in the CCMC CME Scoreboard (https://kauai.ccmc.gsfc.nasa.gov/CMEscoreboard/). In columns 16 and 17, we refer to the results forecasted by the Space Weather Research Center at NASA Goddard Space Flight Center (GSFC/SWRC). For the IPS based forecasting, we selected the simulation run that best estimates the IPS observations. Therefore, if the CME with the initial speed estimated from LASCO or CACTus shows the best match to the IPS, IPS and LASCO or CACTus based forecasting will exhibit the same arrival time. For example, the CACTus based simulation show the best match to the IPS observation among 10 simulation runs of the CME event on April 2, 2014. Therefore, the IPS based forecasting selected the simulation run with initial speed derived from CACTus. In this event, the LASCO based simulation resulted more than 13 h arrival-time error, while the IPS based forecasting, which selected the CACTus based simulation, produced only about 5 h arrival-time error. For the CME event on September 4, 2017, both CACTus and LASCO based simulations estimated IPS distributions that were significantly different from the observed distribution. Hence, the IPS based forecasting selected the different simulation run. That produced a smaller arrival-time error than those derived from both CACTus and LASCO simulations.

**Arrival-time error**

Figure 3a shows a histogram of the average absolute arrival-time errors. The average arrival-time error for CMEs with the initial speed derived from CACTus, LASCO, and the best fit to IPS is 6.7 h, 6.7 h, and 5.0 h, respectively. The corresponding average absolute arrival-time error for the WSA-Enlil-cone model archived in the CCMC CME Scoreboard is 11.9 h. All simulation approaches resulted in earlier average arrival-time prediction, i.e. the average of predicted arrival times is earlier than the average of actual arrival times.

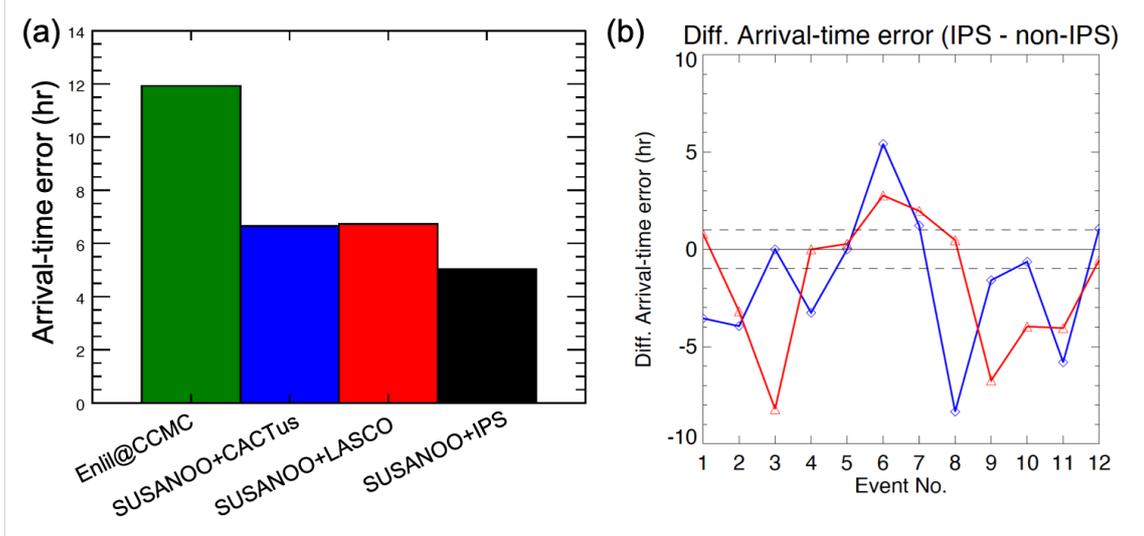

**Figure 3**. **Arrival-time error of CME at the Earth.** (a) Average of the absolute arrival-time error at the Earth forecast by the real-time forecast result by GSFC/SWRC archived in the CCMC website (green), and SUSANOO using different CME initial speeds; with initial speeds derived from CACTus (blue), the LASCO catalog (red), and the best fit to IPS data (black). (b) Difference of the absolute arrival-time error between the SUSANOO simulation that best fit to IPS data and SUSANOO simulation using the initial speeds derived from CACTus (blue) and LASCO (red). The event numbers correspond to the numbers in the first column of Table 1. Dotted lines indicate a 1 h difference, which is equivalent to the error bar of the arrival time in this forecasting method.

**Longitudinal and initial speed dependence of the arrival-time error**

Figure 4a shows the relationship between the predicted arrival-time error and the initial speed of the CMEs. Pearson's correlation coefficient (CC), $R_p$, is 0.29, 0.61, and −0.26, and the Spearman's rank CC, $R_s$, is 0.14, 0.50, and −0.29 with the probability ($p$-value) of the observed (or more extreme) Spearman's rank CC occurring by chance of 0.66, 0.09, and 0.37, with the initial speeds derived from CACTus, LASCO catalog, and the best fit to IPS data, respectively. A weak correlation exists between the arrival-time error and initial speed in white-light based simulations, while no relationship exists in the IPS-based prediction. Figure 4b shows the relationship between the predicted arrival-time error and the longitude of the CME location. Although no linear correlation exists between them, the arrival-time error seems to be the lowest at 20–30 degree for all methods. Although it might be related to the characteristics of our model simulation and/or the observed white light and radio emissions, investigating the reason for this trend is beyond the scope of this study.

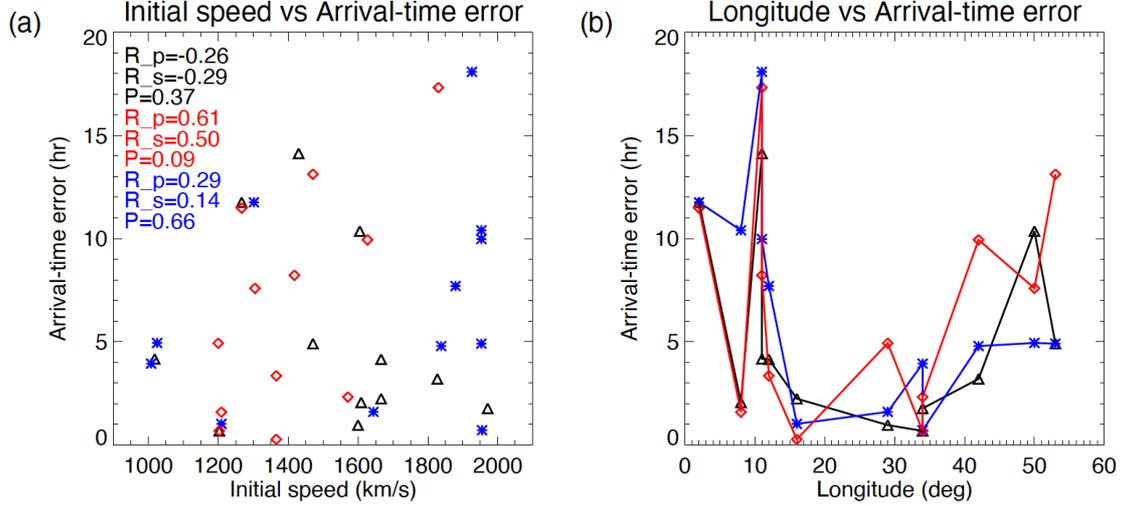

**Figure 4. Initial speed and longitudinal dependence of the arrival-time error.** (a) Relationship between predicted arrival-time error and initial speed of the CMEs. Pearson's CC ($R_p$), Spearman's rank CC ($R_s$), and the P-value (P) are shown in the left top. (b) Relationship between predicted arrival-time error and longitude of CME location. Colors indicate CMEs with initial speeds derived from CACTus (blue), the LASCO catalog (red), and the best fit to IPS data (black).

## Summary and Discussion

**Differences in the arrival-time error**

As shown in Figure 2, the simulated solar wind speed shows a smooth change compared to the actual shocks. The simulation can capture a shock structure with about 3 grid points. If the simulation has a sufficiently high spatial resolution, the shock location could converge to the center of the gradient of the apparent width. However, we set the spatial resolution of the simulation to be as low as possible in order to execute multiple simulation cases within a reasonable lead time from the arrival of the CME by using a general workstation. Because the grid intervals in the radial direction of around 1 AU are approximately 2.3 Rs, the width of a radially propagating shock in the simulation becomes about 7 Rs. It takes approximately 2 hours for the shock of a typical speed 680 km/s to pass through the Earth's position. If a shock structure inclines in the non-radial direction, the apparent shock width will be larger than the width observed for a shock structure that is perpendicular to the radial direction. In the case shown in Figure 2a, the shock plane was oriented about 40–50° from the Sun-Earth line, and therefore, the shock width along the Sun-Earth line becomes about 5 grid points (see 0.95–1.0 AU).

This study used the IPS data to optimize the simulations. The IPS observation detects the density enhancement in front of the CMEs. As described in the previous paragraph, because of the coarse spatial resolution, the simulated CMEs have density distributions that are gradual compared to the actual ones. This study would select the most realistic case by comparing the observed IPS data with the simulated IPS calculated from the gradual density distribution. This means that the front part of the density enhancement of the selected simulation should be the most reliable location for the CME front. Therefore, the onset time of the shock should be defined as the arrival time regardless of the origin of the shock width of the simulations.

IPS data improved the arrival time forecast of the SUSANOO MHD simulation compared to that derived from the SUSANOO simulation without IPS data by 1.7 h (25%). The error bar of the arrival-time in each simulation run is mainly given by the intervals of initial CME speed (100 km/s), which corresponds to about 1h difference in the arrival-time as shown in Figure 2b. Therefore, we can distinguish a difference of arrival-times only larger than 1h. In 10 of the 12 events, IPS-based SUSANOO provided a better or similar (i.e. the difference is less than 1h) arrival time compared to the SUSANOO without IPS, and the arrival time of IPS-based SUSANOO is worse than that of SUSANOO without IPS in only two of the 12 cases as shown in Figure 3b. Therefore, we consider that this improvement from SUSANOO without IPS to IPS-based SUSANOO is encouraging although we need for further study with a much larger sample.

Another point is that the IPS-based SUSANOO significantly improved the arrival-time forecast compared to the SUSANOO without IPS in three CME events No. 3, 8, and 9: (2014-04-02, 2015-06-22, and 2015-06-25), which resulted in large arrival-time errors when forecast by the SUSANOO simulation without IPS (i.e. with either CACTus or LASCO). For example, for the CME event observed on June 25, 2015, the LASCO based simulation produced an arrival-time error of approximately 10 h, while the IPS based forecast selected a different simulation run that produced an arrival-time error of only 3 h approximately. From the perspective of space weather forecasting, a large arrival-time error can have serious consequences, which suggests that IPS data play a crucial role in the forecasting system.

The approach employed in this study provided better arrival-time forecasts than the real-time operated WSA-ENLIL-cone model, even without IPS data, probably because only a limited number of CME events were tested, and this result might not be maintained for a larger data set. However, the probability of our simulation providing a reliable forecast is high. Our simulation used a spheromak instead of a cone as the CME model, resulting in a more realistic reconstruction of CME propagation.

Moreover, the magnetic flux, radial width, and longitudinal width of the spheromak were assumed to be proportional to the X-ray flux of the corresponding flare (see Methods section). If a fixed magnetic flux, radial width, and longitudinal width for all spheromak had been set, the average arrival-time would have been worse. Therefore, this assumption can improve the estimated arrival-time error. This study based on a full 3D simulation that includes the polar regions of the heliosphere that can provide a better reconstruction of the interaction between CMEs and solar wind.

Our simulation approach resulted in earlier arrival-time predictions, which is consistent with previous MHD simulations (Wold et al. 2018). If the magnetic field flux of the spheromak is too strong, acceleration and expansion of the CME can be overestimated, which may result in an earlier prediction. If the size of the CME is too small, the drag force from the background solar wind can be underestimated, which can also result in an earlier prediction. Furthermore, a radio scattering region may exist in front of the CME (Gothoskar and Rao, 1999), which could result in early prediction even when using IPS data.

**Causes of the arrival-time error**
The simulations that best fit the IPS data still exhibited arrival-time errors. In this study, we only optimized the initial speed of the CME; the other parameters were approximately assumed from GOES and LASCO observations. These assumptions should be optimized in future studies. Some CME parameters assumed in this study can be derived from LASCO white-light images by fitting the CME geometry using models such as the graduated cylindrical shell model (e.g., Thernisien et al. 2006). CMEs propagating in interplanetary space can be derived from the Heliospheric Imager onboard STEREO satellites (Möstl et al 2014; Howard 2015). White-light coronagraph data can also be estimated from our MHD simulations by calculating the Thomson scattering along the line of sight. Therefore, we may be able to optimize the CME parameters other than the initial speed by combining IPS observations, white-light observations, and MHD simulations. In addition, the CME model can be improved from the spheromak to some other more realistic models.

The velocity, density, and temperature of background solar wind in our simulation were obtained from empirical models. These models have ambiguities and result in CME arrival-time errors because the interactions between background solar wind and CMEs can affect the propagation of CMEs (e.g., Chen 1996; Vršnak and Gopalswamy 2002). Background solar wind velocity can also be derived from IPS observations using the tomography technique (Kojima et al. 1998; Jackson et al 1998), and the derived velocity distribution can be adopted as the inner boundary of the global MHD simulation (Jackson et al. 2015). In future modeling efforts, we may first derive background solar wind velocity distribution from IPS data using the tomography technique, in which transient phenomena such as CMEs are less prominent because the tomography of the background solar wind requires at least several days of IPS data. Then, we can simulate and fit CMEs using IPS data acquired just before forecasting in which CMEs are more prominent.

**Application to the space weather forecasting system**

The approach employed in this study has been partially installed in the space weather forecasting system of the National Institute of Information and Communications Technology (NICT), which is the Japanese space weather forecasting center. During daily operation, CME arrival times should be forecasted as soon as possible after a CME is observed and the initial forecast should be given automatically or semi-automatically by human forecasters with a large error range. IPS data can be made available approximately 1–2 days after the onset of CMEs exhibiting typical speeds. IPS data cannot be used for the initial forecast itself but can be used to limit the range of the initial forecast, which is similar to data assimilation forecasts of typhoon trajectories. It takes about 2–4 days for the typical CMEs to reach the Earth from their onset. Therefore, our IPS-based forecast will work for most CMEs. The IPS-based forecast may not be able to record the signature of the fastest CME, which is generally the most significant and which arrives at the Earth within 1 day. Coordinated observations of several IPS stations on different longitudes will improve the observation cadence of the IPS, which may lead to further improvement in the accuracy of forecasting the arrival of extremely fast CMEs.

## Declarations

**Ethics approval and consent to participate**
Not applicable

**Consent for publication**
Not applicable

**List of abbreviations**
CCMC: community coordinated modeling center
CACTus: computer aided CME tracking software
CME: coronal mass ejection
GOES: geostationary operational environmental satellite
HI: heliospheric imager
IPS: interplanetary scintillation
LASCO: large angle and spectrometric coronagraph
MHD: magnetohydrodynamic
PFSS: potential field source surface
STEREO: solar terrestrial relations observatory
SOHO: solar and heliospheric observatory

SUSANOO: space-weather-forecast-usable system anchored by numerical operations and observations
SWIFT: solar wind imaging facility
WSA: Wang–Sheeley–Arge

**Availability of data and materials**
The datasets supporting the conclusions of this article are available in the data repository of ISEE, Nagoya University http://stsw1.isee.nagoya-u.ac.jp/ips_data-e.html, Virtual Solar Observatory https://sdac.virtualsolar.org/cgi/search, and OMNIweb service https://omniweb.gsfc.nasa.gov/.

**Competing interests**
The authors declare that they have no competing interests.

**Funding**
This work was supported by the Ministry of Education, Culture, Sports, Science and Technology (MEXT), Japan Society for the Promotion of Science (JSPS), KAKENHI Grant Number 18H04442, 15H05813, and 15H05814.

**Authors' contributions**
KI led this study and drafted the manuscript. DS developed the MHD simulation codes. KI, MT, and KF operated and maintained the IPS radio observations. DS, MD, and YK operated and maintained the space weather forecasting system based on the MHD simulations. All authors read and approved the final manuscript.

**Acknowledgements**
This work was supported by MEXT/JSPS KAKENHI Grant Number 18H04442, 15H05813 and 15H05814. The IPS observations were provided by the solar wind program of the Institute for Space-Earth Environmental Research (ISEE). This study was conducted using the computational resources of the Center for Integrated Data Science, ISEE, Nagoya University, through a joint research program. We thank the LASCO coronagraph group for the white-light CME images.

**Table 1. CME characteristics investigated in this study and their observed and forecasted arrival times.**

| CME | | | | | | CACTus | | | LASCO | | | IPS | | | CCMC (ENLIL) | |
|---|---|---|---|---|---|---|---|---|---|---|---|---|---|---|---|---|
| No | Onset time | X-ray Class | Longitude | Latitude | Arrival time | Arrival time | Speed | abs(diff) | Arrival time | Speed | abs(diff) | Arrival time | Speed | abs(diff) | Arrival time | abs(diff) |
| 1 | 2013/3/15 | X1.2 | -12 | 11 | 2013-03-17T05:28 | 2013/3/16 21:46 | 1879 | 7.70 | 2013/3/17 8:49 | 1366 | 3.35 | 2013/3/17 1:18 | 1666 | 4.15 | 2013-03-16T16:59 | 12.48 |
| 2 | 2014/1/7 | X1.2 | 11 | 15 | 2014-01-09T19:32 | 2014/1/9 1:27 | 1926 | 18.08 | 2014/1/9 2:13 | 1830 | 17.32 | 2014/1/9 5:24 | 1430 | 14.13 | 2014-01-09T00:38 | 18.90 |
| 3 | 2014/4/2 | M6.5 | -53 | 14 | 2014-04-05T09:40 | 2014/4/5 14:34 | 1953 | 4.91 | 2014/4/5 22:47 | 1471 | 13.12 | 2014/4/5 14:34 | 1471 | 4.91 | 2014-04-04T09:21 | 24.32 |
| 4 | 2014/4/18 | M7.3 | 34 | -20 | 2014-04-20T10:22 | 2014/4/20 14:18 | 1008 | 3.94 | 2014/4/20 11:02 | 1203 | 0.68 | 2014/4/20 11:02 | 1203 | 0.68 | 2014-04-20T09:09 | 1.22 |
| 5 | 2014/9/10 | X1.6 | -2 | 14 | 2014-09-12T15:26 | 2014/9/12 3:40 | 1302 | 11.76 | 2014/9/12 3:57 | 1267 | 11.48 | 2014/9/12 3:40 | 1267 | 11.76 | 2014-09-12T11:47 | 3.65 |
| 6 | 2015/6/18 | M3.0 | -50 | 15 | 2015-06-21T15:40 | 2015/6/21 10:43 | 1025 | 4.94 | 2015/6/21 8:04 | 1305 | 7.59 | 2015/6/21 5:18 | 1605 | 10.36 | 2015-06-21T09:26 | 6.23 |
| 7 | 2015/6/21 | M2.6 | -16 | 12 | 2015-06-22T17:59 | 2015/6/22 16:57 | 1209 | 1.02 | 2015/6/22 18:14 | 1366 | 0.27 | 2015/6/22 15:44 | 1666 | 2.24 | 2015-06-22T21:43 | 3.73 |
| 8 | 2015/6/22 | M6.5 | 8 | 12 | 2015-06-24T12:57 | 2015/6/24 2:32 | 1953 | 10.40 | 2015/6/24 14:32 | 1209 | 1.59 | 2015/6/24 10:53 | 1609 | 2.06 | 2015-06-24T18:18 | 5.35 |
| 9 | 2015/6/25 | M7.9 | 42 | 9 | 2015-06-27T03:30 | 2015/6/27 8:17 | 1838 | 4.79 | 2015/6/27 13:26 | 1627 | 9.94 | 2015/6/27 6:42 | 1827 | 3.20 | 2015-06-28T02:00 | 22.50 |
| 10 | 2017/7/14 | M2.4 | 29 | -6 | 2017-07-16T05:14 | 2017/7/16 3:38 | 1644 | 1.60 | 2017/7/16 10:09 | 1200 | 4.93 | 2017/7/16 4:16 | 1600 | 0.96 | 2017-07-16T21:42 | 16.47 |
| 11 | 2017/9/4 | M5.5 | 11 | -10 | 2017-09-06T23:08 | 2017/9/6 13:08 | 1953 | 9.98 | 2017/9/6 14:54 | 1418 | 8.22 | 2017/9/6 18:57 | 1018 | 4.17 | 2017-09-06T14:51 | 8.28 |
| 12 | 2017/9/6 | X9.3 | 34 | -9 | 2017-09-07T22:30 | 2017/9/7 21:47 | 1955 | 0.71 | 2017/9/8 0:49 | 1571 | 2.32 | 2017/9/7 20:44 | 1971 | 1.77 | 2017-09-08T18:27 | 19.95 |